\documentclass[aps, prl,twocolumn]{revtex4-1}
\usepackage{float}
\usepackage{graphicx}
\usepackage{amsmath}
\usepackage{amssymb}
\usepackage{color}
\usepackage{soul}
\usepackage{subfigure}
\usepackage{soul}
\usepackage{braket}
\usepackage[english]{babel}
\newcommand{\E}{\mathcal{E}}
\usepackage[sort&compress]{natbib}
\begin{document}
\title{Coherent anti-Stokes Raman Spectroscopy of single and multi-layer graphene}
\author{%
A. Virga$^{1,2,*}$,
C. Ferrante$^{3,1,*,\dagger}$,
G. Batignani$^{1}$,
D. De Fazio$^{4}$,
A. D. G. Nunn$^{2}$,
A. C. Ferrari$^{4}$,
G. Cerullo$^{5}$,
T. Scopigno$^{1,3,^\ddagger}$}
\affiliation{$^{1}$Dipartimento di Fisica,~Universit\'a~di~Roma~\textquotedblleft La Sapienza",~I-00185,~Roma,~Italy}
\affiliation{$^{2}$Istituto Italiano di Tecnologia, Center for Life Nano Science @Sapienza, Roma, I-00161, Italy}
\affiliation{$^{3}$Istituto Italiano di Tecnologia, Graphene Labs, Via Morego 30, I-16163 Genova, Italy}
\affiliation{$^4$Cambridge Graphene Centre, Cambridge University, 9 JJ Thomson Avenue, Cambridge CB3 OFA, UK}
\affiliation{$^5$IFN-CNR, Dipartimento di Fisica, Politecnico di Milano, P.zza L. da Vinci 32, 20133 Milano, Italy}
\affiliation{$^*$These authors contributed equally to this work}
\affiliation{$^\dagger$carino.ferrante@iit.it}
\affiliation{$^\ddagger$tullio.scopigno@roma1.infn.it}
\begin{abstract}
We report stimulated Raman spectroscopy of the G phonon in both single and multi-layer graphene, through Coherent anti-Stokes Raman Scattering (CARS). The signal generated by the third order nonlinearity is dominated by a vibrationally non-resonant background (NVRB), which obscures the Raman lineshape. We demonstrate that the vibrationally resonant CARS peak can be measured by reducing the temporal overlap of the laser excitation pulses, suppressing the NVRB. We model the observed spectra, taking into account the electronically resonant nature of both CARS and NVRB. We show that CARS can be used for graphene imaging with vibrational sensitivity.
\end{abstract}
\maketitle
\begin{figure}
\centerline{\includegraphics[width=90mm]{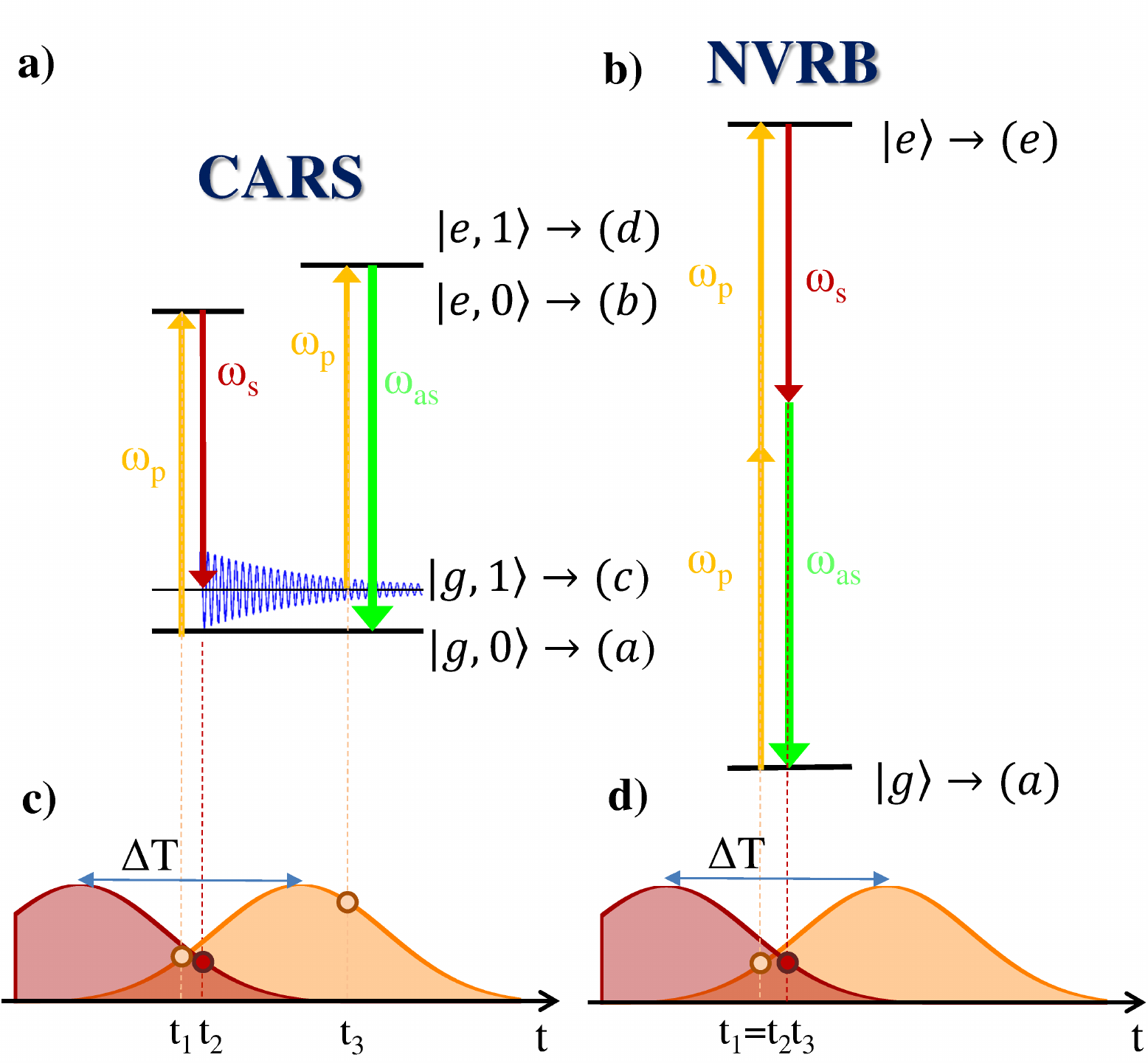}}
\caption{Schematic of CARS and NRVB third-order nonlinear processes. Interaction with pulses $\omega_P$, $\omega_S$, results in blue-shifted a) CARS and b) NVRB contributions at $\omega_{as}=2\omega_P-\omega_S$. Since in CARS a vibrational coherence is stimulated by two consecutive interactions with the pump and Stokes fields, their frequency difference must correspond to a Raman active mode, $\omega_P-\omega_S=\omega_v$. c,d) Constraints for the temporal sequence of the field-matter interactions (represented by circles at the top of the pulse envelopes), for CARS and NVRB. In NVRB, the 3 interactions generating the $\chi^{(3)}$ signal happen within the few fs electronic dephasing time\cite{Mukamel1999}. In CARS, the third interaction can occur over the much longer vibrational dephasing time (a few ps)\cite{Mukamel1999}, within the pump pulse (PP) temporal envelope.}
\label{fig1}
\end{figure}
Single-layer graphene (SLG) has a high nonlinear third-order susceptibility: $\lvert\chi^{(3)}\rvert$ $\sim10^{-10}$ e.s.u for harmonic generation\cite{Soavi2018} and $\lvert\chi^{(3)}\rvert$ $\sim10^{-7}$ e.s.u. for frequency mixing\cite{Hendry2010}, where one electrostatic unit of charge (1 e.s.u), in standard units (SI) is\cite{BoydBook}: $\chi^{(3)}$(SI)/$\chi^{(3)}$(e.s.u.) = $4\pi/(3\times 10^{4})^2$. This is up to seven orders of magnitude greater than those of dielectric materials such as silica ($\chi^{(3)} = 1.4 \times 10^{-14}$ e.s.u\cite{PotmaMukamel2013}). This property is due to optical resonance with interband electronic transitions\cite{Khurgin2014} and has led to the observation of gate-tunable third-harmonic generation\cite{Soavi2018} and nonlinear four-wave mixing\cite{Hendry2010,Wu2011,Gu2012} (FWM, i.e. the third-order processes whereby an electromagnetic field is emitted by the nonlinear polarization induced by three field-matter interactions\cite{BoydBook}). FWM can be exploited for for graphene imaging, with an image contrast of up to seven orders of magnitude\cite{Hendry2010} higher than that of optical reflection microscopy\cite{duong_probing_2012}. However, FWM-based imaging reported to date in graphene\cite{Hendry2010} lacks chemical selectivity and does not provide the same wealth of information brought about by the vibrational sensitivity of Raman spectroscopy\cite{FerrariPRL2006, FerrariReview2013}.

Coherent anti-Stokes Raman scattering (CARS)\cite{Begley1974,Duncan1982,Zumbusch1999,Cheng2002} is a FWM process that exploits the nonlinear interaction of two laser beams, the pump field $E_P$ at frequency $\omega_P$ and the Stokes field $E_S$ at frequency $\omega_S<\omega_P$, to access the vibrational properties of a material. As shown in Fig.\ref{fig1}a, when the energy difference between the two photons matches a phonon energy ($\hbar \omega_P-\hbar\omega_S=\hbar\omega_v$), the interaction of the laser pulses and the sample results in the generation of vibrational coherences\cite{PotmaMukamel2013}. While spontaneous Raman (SR) scattering is an incoherent signal\cite{Mukamel1999}, since the phases of the electromagnetic fields emitted by individual scatterers are uncorrelated\cite{Mukamel1999}, in CARS, atomic vibrations are coherently stimulated, i.e. atoms oscillate with the same phase\cite{PotmaMukamel2013}, potentially leading to a signal enhancement of several orders of magnitude depending on incident power and scatterer density\cite{Cui2009,Pestov265}.

The same combination of optical fields used for CARS can generate another FWM signal, a non-vibrationally resonant background (NVRB)\cite{Hendry2010}, Fig.\ref{fig1}b. In both processes, the optical response consists of a field emitted at the anti-Stokes frequency $\omega_{as}=2\omega_P-\omega_s$\cite{PotmaMukamel2013}. However, the interference of the two effects usually generates an an additional contribution which is dispersive with respect to the emitted optical frequency, i.e. shaped as the first derivative of a peaked function (resembling the real part of the refractive index around a resonance), which introduces an asymmetric distortion of the Raman peak profile in the region $\omega_{as}=\omega_P+\omega_v$ \cite{evans_coherent_2008}.

In the biological field\cite{Cui2009,Evans16807}, a wealth of studies has demonstrated the potential of CARS for fast imaging\cite{Cui2009, Pestov265, CiceroneNat2015}, with pixel acquisition times as low as$\sim0.16\mu s$\cite{Evans16807}, thus allowing for video-rate microscopy\cite{Evans16807}. By contrast, there are only a few reports to date of CARS imaging of micro-structured materials (such as polyethylene blend\cite{Lee2012}, multicomponent polymers\cite{Lim2006}, cholesterol micro-crystals\cite{Lim2011}) and nano-structured ones (patterned gold surfaces\cite{Steuwe2011}, single wall nanotubes\cite{Ikeda2009,Lee2010}, highly oriented pyrolytic graphite\cite{Dovbeshko2010}). Such studies, performed with pixel acquisition times down to$\sim2\mu s$\cite{Baldacchini2010}, have shown the ability of CARS to identify chemical heterogeneities on sub-micrometer scales and characterize single particles that are part of a larger domain, enabling e.g. to visualize microscopic domains (polystyrene, poly-methyl methacrylate, and poly-ethylene terephthalate) in the case of the above mentioned polymer mixtures\cite{Potma2013}, or to provide detailed maps of microcrystal orientation in organic matrices (e.g. cholesterols in atherosclerotic plaques\cite{Lim2011}).

In graphene, despite the large $\chi^{(3)}$\cite{Hendry2010,Soavi2018}, no CARS peak profiles, equivalent to those measured in SR, have been observed to date, to the best of our knowledge. We reported SR with single-color pulsed excitation\cite{ferrante_nc}, using the same picosecond lasers usually adopted for CARS\cite{Evans16807}. However, in order to measure CARS, a combination of pulses with different colors must be adopted\cite{TRCARS2017}. By scanning the pulse frequency detuning in a two-color experiment, a dip has been observed in the third order nonlinear spectral response of SLG at the G phonon frequency. This was interpreted as an anomalous antiresonance and phenomenologically described in terms of a Fano lineshape\cite{LafetaArx2017}.

\begin{figure}
\centerline{\includegraphics[width=90mm]{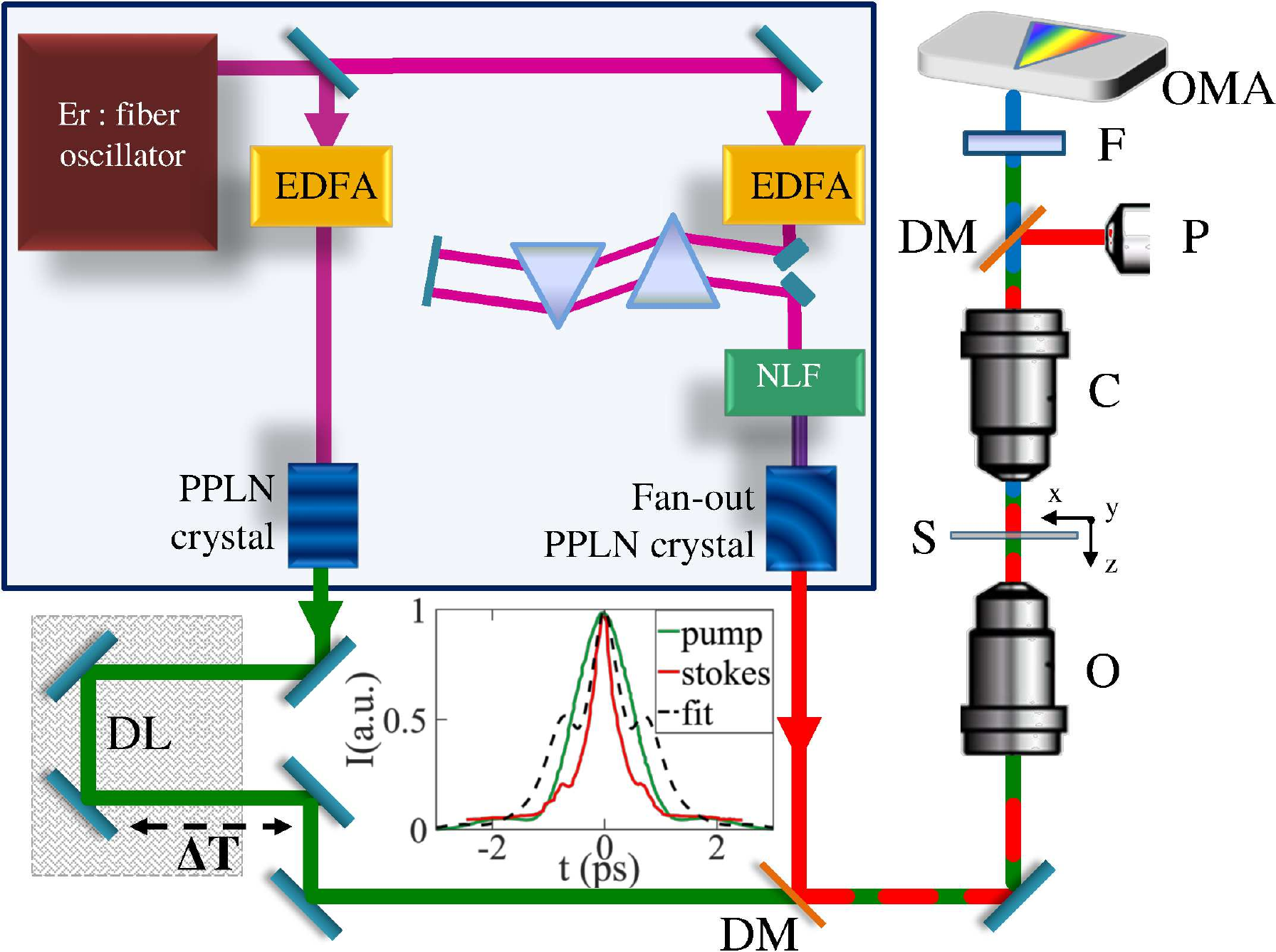}}
\caption{CARS setup. EDFA, erbium-doped fiber amplified; NLF, nonlinear fiber for SC generation; DL, delay line; DM, dichroic mirror; O, objective; S, sample; C, condenser; P, powermeter; F, filter; OMA, optical multichannel analyzer. Purple, green, and red lines represent the beam pathways of 1550nm, 784 nm (PP) and  tunable SP. The second harmonic autocorrelation of PP (green line) and SP (red line) are reported in the inset. The black dashed line simulates the autocorrelation obtained by using the profile from the best fit (colored dashed lines) in Fig.\ref{fig2}.}
\label{figsetup}
\end{figure}

Here we use two 1ps pulses (see inset of Fig.\ref{figsetup}) to explore FWM in SLG and few-layer graphene (FLG). We experimentally demonstrate and theoretically describe how the inter-pulse delay, $\Delta T$ (Figs.\ref{fig1}c-d) can be used to modify the relative weight of CARS and NVRB components that simultaneously contribute to the FWM, thus recovering the G-band Raman peak profile.
We show that the dip in the nonlinear optical response around the vibrational resonance is due to the interplay of CARS and NVRB under electronically resonant conditions, which allows vibrational imaging with signal levels as large as those of the third-order nonlinear response.

SLG is grown on a 35$\mu$m copper (Cu) foil, following the process described in Ref.\citenum{LiS2009}. The substrate is heated up to 1000$^{\circ}$C and annealed in hydrogen atmosphere (H$_2$, 20 sccm) for 30 minutes at $\sim$200 mTorr. Then, 5 sccm of methane (CH$_4$) are let into the chamber for the following 30 minutes to enable growth\cite{BaeNN2010,LiS2009}. The sample is then cooled back to room temperature in vacuum ($\sim$1 mTorr) and unloaded from the chamber. SLG is subsequently transferred on a glass substrate by a wet method. Poly-methyl methacrylate (PMMA) is spin coated on the SLG/Cu and floated on a solution of ammonium persulfate (APS) and deionized water. When Cu is etched\cite{BaeNN2010,BonaMT2012}, the PMMA membrane with attached SLG is then moved to a beaker with deionized water for cleaning APS residuals. The membrane is subsequently lifted with the target substrate. After drying, PMMA is removed in acetone leaving SLG on glass. SLG is characterized by SR after transfer using a Renishaw InVia spectrometer at 514nm. The position of the G peak, Pos(G), is$\sim$1582cm$^{-1}$, while FWHM(G)$\sim$14cm$^{-1}$. The 2D to G peak area ratio is$\sim$5.3, indicating a p-doping after transfer$\sim$250meV\cite{BaskPRB2009, DasNN2008}, which corresponds to a carrier concentration$\sim$5$\cdot$10$^{12}$cm$^{-2}$. FLG flakes are produced by micromechanical cleavage from bulk graphite\cite{Novoselov2005PNAS}. The bulk crystal is exfoliated on Nitto Denko tape. The FLG G peak is$\sim$1580cm$^{-1}$. The D peak is negligible. The 2D peak shape indicates this is Bernal stacked FLG\cite{FerrariPRL2006,FerrariReview2013}.
\begin{figure}
\centerline{\includegraphics[width=90mm]{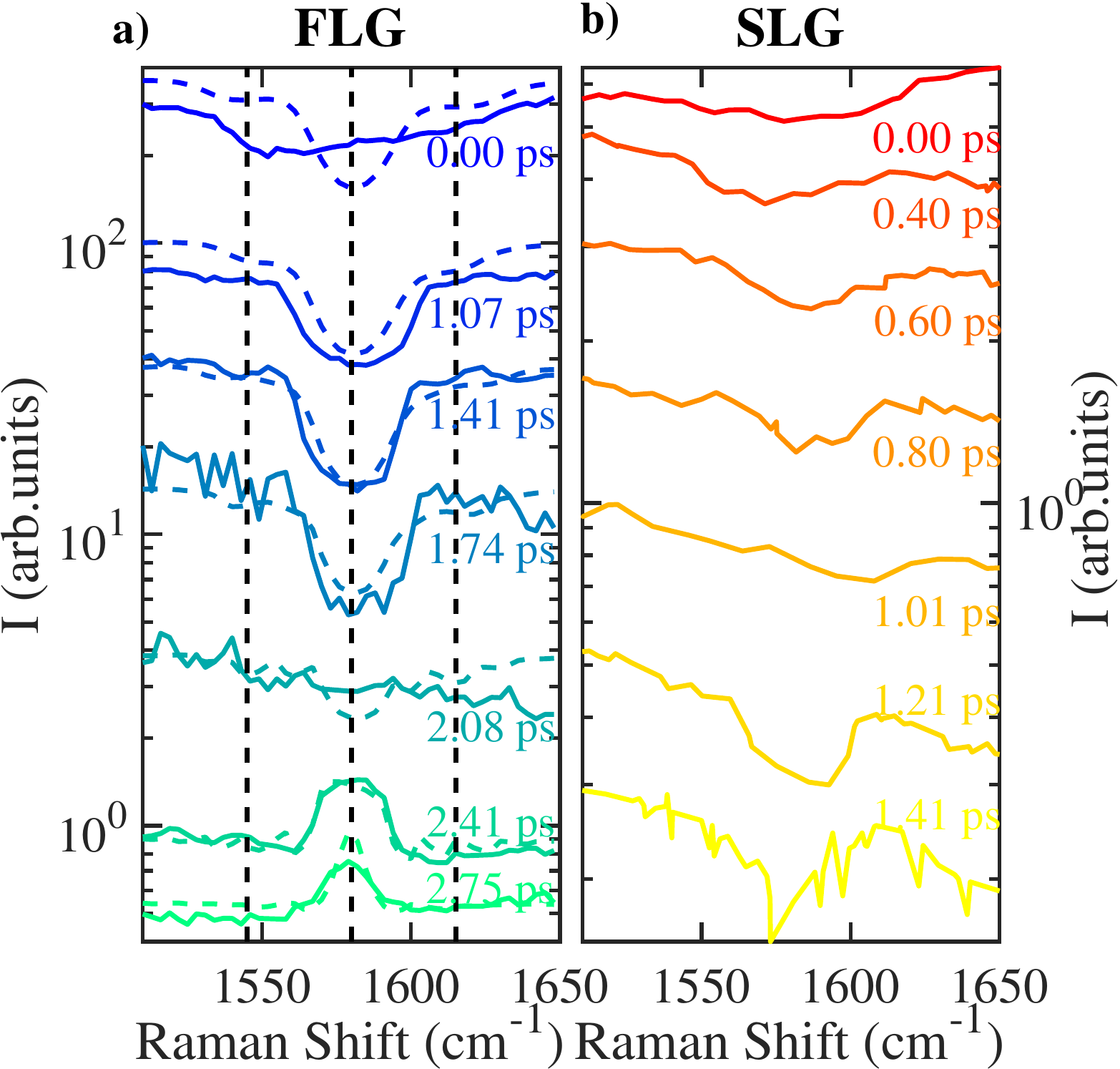}}
\caption{CARS spectra of (a) FLG and (b) SLG  as a function of Raman shift ($\tilde{\nu}-\tilde{\nu}_P$) at different $\Delta T$ between the beams at tunable $\omega_S$ and fixed $\omega_P$. In (a), colored dashed lines are fits to the data using Eq.\ref{eq:FWMgeneral} and the nonlinear polarization obtained from Eqs.\ref{eq:simulation2}-\ref{eq:simulationNVRB2}. Vertical black dashed lines indicate three energies ($\tilde{\nu}_{1,2,3}-\tilde{\nu}_P=$ 1545, 1576, 1607 cm$^{-1}$), taken as reference for the FLG CARS images in Fig.\ref{fig3}.}
\label{fig2}
\end{figure}

For CARS experiments, we use a two-modules Toptica FemtoFiber Pro source, with two Erbium-Doped fiber amplifiers (EDFA) at$\sim1550$nm generating 90fs pulses at 40MHz, seeded by a common mode-locked Er:fiber oscillator\cite{Krauss:09}, Fig.\ref{figsetup}. In the first branch (FemtoFiber pro NIR), 1ps pulses at 784nm (pump pulse, PP) are produced by second-harmonic spectral compression\cite{Marangoni:09} in a 1cm Periodically Poled Lithium Niobate (PPLN) crystal. In the second branch (FemtoFiber pro TNIR), the amplified laser passes through a nonlinear fiber (NLF), wherein a supercontinuum (SC) output is generated. The SC spectral intensity can be tuned with a motorized Si-prism-pair compressor. A PPLN crystal with a fan-out grating (a poling period changing along the transverse direction) is exploited to produce broadly tunable (from 840 to 1100nm) narrowband 1ps Stokes pulses (SP), with a power$<10$mW. A dichroic mirror (DM) is used to combine the two beams, whose relative temporal delay is tuned with an optical delay line (DL). A long-working distance 20x objective (O, numerical aperture NA=0.4) focuses the pulses onto the sample (S). The generated and transmitted light is collected by a condenser (C) and the PP and SP are filtered out by a short-pass filter (F). The total FWM signal is collected with an optical multichannel analyser (OMA, Photon Control SPM-002-E). A dichroic mirror reflects the SP in order to measure its intensity ($I_s$) with a powermeter (P). FWM spectra are obtained by scanning $\omega_S$ around $\omega_P-\omega_v$ (at fixed $\omega_P$) to probe the G-band phonon frequency, $\omega_v=\omega_G$. Fig.\ref{fig2} displays the FWM intensity, normalized to $I_s$, for different $\Delta T$.

In both SLG and FLG measurements, for $\Delta T$ shorter than the vibrational dephasing time $\tau\sim$1ps\cite{Bonini2007}, i.e. the characteristic time of coherence loss\cite{Mukamel1999}, a Lorentzian dip at $\omega_{as}=\omega_P+\omega_G$ appears on top of a background\cite{LafetaArx2017}. For $\Delta T>2$ps, while the total FWM signal decreases by nearly two orders of magnitude, the dip observed in FLG at $\Delta T\sim0$ps evolves into a Raman peak shape at the G-phonon energy. No dispersive features are observed at any $\Delta T$, unlike what normally expected for the interference between NVRB and CARS\cite{evans_coherent_2008}. Here we use pulses with duration $\delta t\sim1$ps since this allows us to scan the inter-pulse delay across the vibrational dephasing time $\tau$ to suppress the NVRB cross section more than the vibrational contribution, while minimizing the spectral broadening due to the finite pulse duration $1/\delta t\simeq$15cm$^{-1}$ \cite{Mukamel1999}.

Since both CARS and NVRB signals depend quadratically on the number of scatterers\cite{Potma2013}, the SLG signal intensity is significantly reduced with respect to FLG (Fig.\ref{fig2}), with a lower signal-to-noise ratio, hampering the detection of peak-shaped vibrational resonances expected for $\Delta T>1.4$ps.

\begin{figure*}
\centerline{\includegraphics[width=180mm]{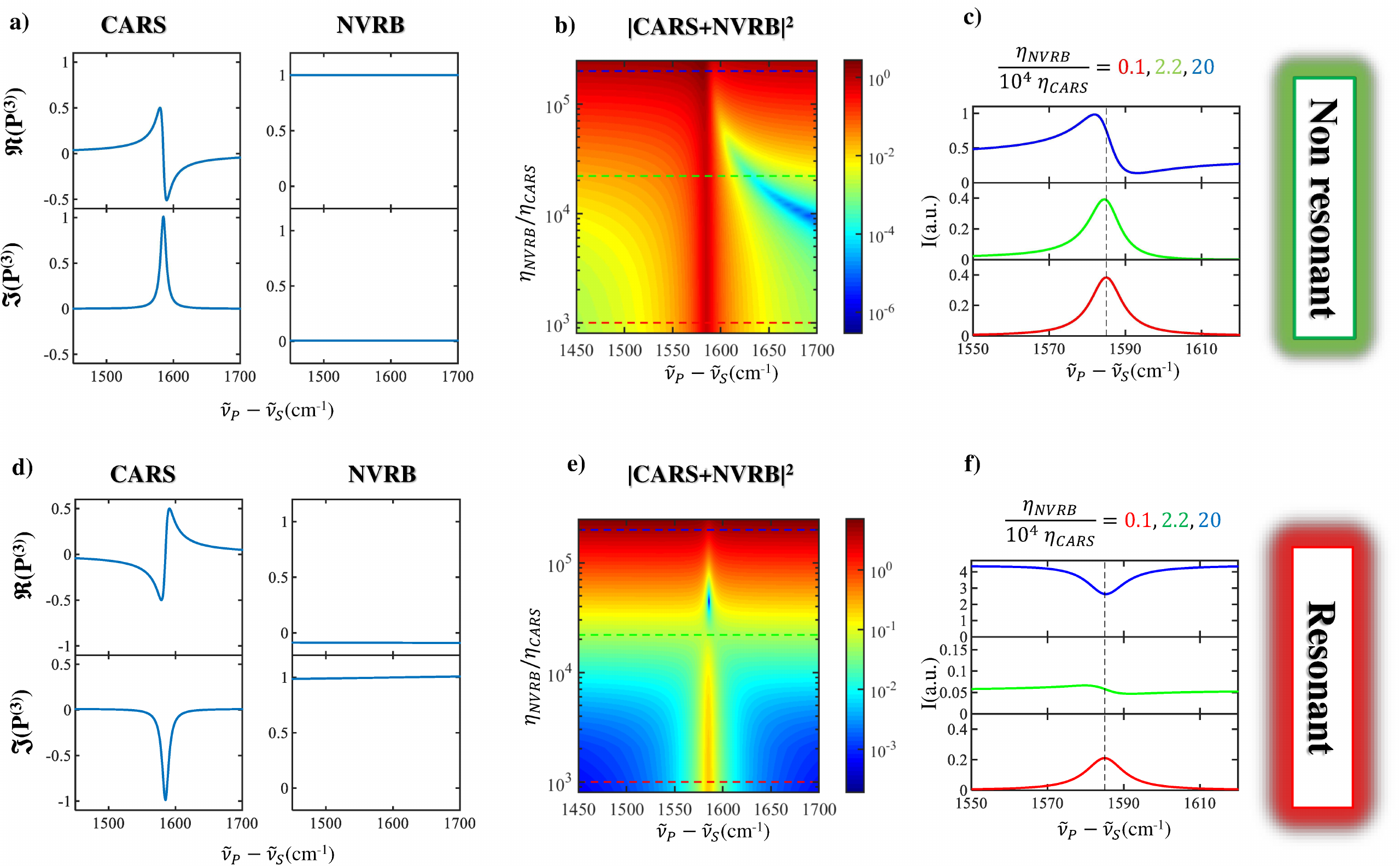}}
\caption{CARS and NVRB spectral profiles for (a,b,c) electronically non-resonant and (d,e,f) resonant regimes, as derived from Eqs.\ref{eq:Pdelta_cars},\ref{eq:Pdelta_nvrb}, considering  $\tau_{ba}=\tau_{da}=\tau_{ea}=10$fs\cite{Brida2013}, $\gamma_{ca}=FWHM(G)/2=6$cm$^{-1}$\cite{Bonini2007}. (a,c) Normalized $\Re(P^{(3)}_{CARS})$, $\Im(P^{(3)}_{CARS})$ and $\Re(P^{(3)}_{NVRB})$, $\Im(P^{(3)}_{NVRB})$. Colormaps in (b,e) generalize (a,c) for different $\eta_{\hbox{\footnotesize{NVRB}}}/\eta_{\hbox{\footnotesize{CARS}}}$, as for Eq.\ref{eq:FWMinterf} and \ref{eq:FWMinterf_Resonant}. In (c,e), selected spectra corresponding to three $\eta_{\hbox{\footnotesize{NVRB}}}/\eta_{\hbox{\footnotesize{CARS}}}$ from the colormap are reported.}
\label{fig1_bis}
\end{figure*}

The data in Fig.\ref{fig2} can be qualitatively understood as follows. The anti-Stokes signal, $I(\omega_{as})$, generated by CARS and NVRB, is proportional to the square modulus of the electric field emitted by the third-order polarization, $P^{(3)}$\cite{PotmaMukamel2013}, as:
\begin{equation}
I(\omega_{as})\propto \lvert P^{(3)}_{CARS}(\omega_{as})+P^{(3)}_{NVRB}(\omega_{as})\rvert^2.\\
\label{eq:FWMgeneral}
\end{equation}
CARS and NVRB signals are simultaneously generated by two light-matter interactions with the PP and a single interaction with the SP, with different time ordering, Figs.\ref{fig1}a-b. Consequently, $P^{(3)}\propto E_P^2 E_S^*$, where * indicates the complex conjugate. However, the temporal constraints for such interactions are significantly different for the two cases. As shown in Figs.\ref{fig1}c-d, in the case of NVRB the three interactions must take place within the dephasing time of the involved electronic excitation, which in SLG is$\sim$10fs\cite{Tomadin2013,Brida2013}, i.e. much shorter than the pulses duration ($\delta t\sim$1ps). Hence, $P^{(3)}_{NVRB}(\omega_{as})$ is only generated during the temporal overlap between the two pulses $P^{(3)}_{NVRB}\propto E_P^2(t-\Delta T)E_S^*(t)$ (the three field interactions, in a representative NVRB event, are indicated by three nearly coincident dots in Fig.\ref{fig1}d). In CARS, the electronic dephasing time only constrains the lag between the first two interactions that generate the vibrational coherence (the two stimulating-field interactions are represented by the two nearly coincident dots in Fig.\ref{fig1}c). This can be read out by the third field interaction within the phonon dephasing time, $\tau\sim 1$ps\cite{Bonini2007} (indicated, for a representative CARS event, by the third dot in Fig.\ref{fig1}d). Thus, $P^{(3)}_{CARS}\propto E_P(t-\Delta T)E_S^*(t)\int_{-\infty}^t E_P(t'-\Delta T) e^{-t'/\tau}dt'$\cite{Volkmer2002}. Therefore, $\Delta T$ can be used to control the relative weights of $P^{(3)}_{CARS}$ and $P^{(3)}_{NVRB}$\cite{Kamga:80, Volkmer2002, Sidorov-Biryukov:06, Pestov265, Selm:10, Kumar:11}. For positive time delays within a few $\tau$, $P^{(3)}_{CARS}/P^{(3)}_{NVRB}$ is progressively enhanced, as shown in Fig.\ref{fig1_bis}a,b,c.

The system response can be evaluated through a density-matrix description of $P^{(3)}(\omega,\Delta T)$\cite{Mukamel1999}. In the presence of a temporal delay between PP and SP, their electric fields can be written as\cite{BoydBook}: $E_{P}(t,\Delta T)={A}_{P}(t,\Delta T)e^{-i\omega_P t}$ and $E_{S}(t,0)={A}_{S}(t,0)e^{-i\omega_S t}$, where ${A}_{P/S}(t,\Delta T)$ indicates the PP/SP temporal envelope. By Fourier transform, the fields can be expressed in the frequency domain as: $\hat{E}_{P}(\omega,\Delta T)= \int_{-\infty}^{+\infty}{E}_{P}(t,\Delta T) e^{i\omega t} dt$ and $\hat{E}_{S}(\omega,0)= \int_{-\infty}^{+\infty}{E}_{S}(t,0) e^{i\omega t} dt$, which can be used to calculate $P^{(3)}_{CARS}(\omega,\Delta T)$ as\cite{Mukamel1999,Batignani2016}:
\begin{widetext}
	\begin{multline}
	P^{(3)}_{CARS}(\omega,\Delta T) \propto -
	\eta_{\hbox{\footnotesize{CARS}}}
	\\  \int_{-\infty}^{\infty}d\omega_3 \int_{-\infty}^{\infty}d\omega_2 \int_{-\infty}^{\infty}d\omega_1
	\frac{\hat{A}_P(\omega_3,\Delta T)\hat{A}_P(\omega_1,\Delta T)\hat{A}_S^*(\omega_2,0)\delta(\omega-2\omega_P+\omega_S-\omega_3-\omega_1+\omega_2)}{\left(\omega_P+\omega_3-\bar{\omega}_{ba}\right)\left(\omega_P-\omega_S+\omega_3-\omega_2-\bar{\omega}_{ca}\right)\left(2\omega_P-\omega_S+\omega_3-\omega_2+\omega_1-\bar{\omega}_{da}\right)}
	\label{eq:CARS}
	\end{multline}
\end{widetext}
where $\eta_{\hbox{\footnotesize{CARS}}}= n_{\hbox{\footnotesize{CARS}}} \mu_{ba} \mu_{cb} \mu_{cd} \mu_{ad}$, $\mu_{ij}$ is the transition dipole moment between the $i$ and $j$ states, $n_{\hbox{\footnotesize{CARS}}}$ is the number of scatterers involved in the CARS process, $\bar{\omega}_{ij}=\omega_{ij}-i\gamma_{ij}=\omega_{i}-\omega_j-i\gamma_{ij}$, $\omega_{ij}=\omega_{i}-\omega_j$ is the energy difference between the levels i and j, and $\gamma_{ij}=\tau_{ij}^{-1}$ is the  dephasing rate of the $\ket{i}\bra{j}$ coherence\cite{Mukamel1999}; $a$ and $c$ denote the vibrational ground state $|g,0\rangle$, and the first vibrational excited level, $|g,1\rangle$, with respect to the electronic ground state $|g\rangle$ ($\pi$ band). In our experiments, $c$ corresponds to the G phonon at $q\sim0$, $b$ and $d$ indicate the vibrational ground state, $|e,0\rangle$, and the first vibrational excited level, $|e,1\rangle$, with respect to the excited electronic state $|e\rangle$ ($\pi^*$ band).

Using the conservation of energy represented by the $\delta$-distribution in Eq.\ref{eq:CARS} and integrating over $\omega_2$, we get:
\begin{multline}\label{eq:simulation}
P^{(3)}_{CARS}(\omega,\Delta T) \propto
-\eta_{\hbox{\footnotesize{CARS}}}
\int_{-\infty}^{\infty}d\omega_1
\int_{-\infty}^{\infty}d\omega_3\\
\frac{\hat{A}_P(\omega_3,\Delta T)\hat{A}_P(\omega_1,\Delta T)\hat{A}_S^*(2\omega_P-\omega_S-\omega+\omega_3+\omega_1,0)}
{\left(
	\omega_P+\omega_3-\bar{\omega}_{ba}
	\right)
	\left(
	\omega-\omega_P-\omega_1-\bar{\omega}_{ca}
	\right)
	\left(
	\omega-\bar{\omega}_{da}
	\right)}
\end{multline}
Defining $\tilde{\nu}=\omega/(2\pi c)$, the third-order nonlinear polarization can be expressed as a function of the Raman shift ($\tilde{\nu}-\tilde{\nu}_P$) as $P^{(3)}(\omega,\Delta T)=P^{(3)}(2\pi c \tilde{\nu},\Delta T)$.

The ${\omega}_{ca}$ level in the denominator of Eq.\ref{eq:simulation} is the frequency of the Raman mode coherently stimulated in CARS, while ${\omega}_{ba}$ and ${\omega}_{da}$ are frequency differences between the electronic levels. In the case of real levels, resonance enhancement occurs\cite{Mukamel1999}. In view of the optical nature of the involved phonons ($q\sim0$), and due to momentum conservation, only one electronic level must be included in the calculation and, consequently, the nonlinear response can be derived for $\omega_{ba}=\omega_{dc}=\omega_P$. In a similar manner, $P^{(3)}_{NVRB}$ can be expressed as\cite{Mukamel1999}:
\begin{multline}\label{eq:simulationNVRB}
P^{(3)}_{NVRB}(\omega,\Delta T) \propto
-\eta_{\hbox{\footnotesize{NVRB}}}
\int_{-\infty}^{\infty}d\omega_1
\int_{-\infty}^{\infty}d\omega_2\\
\frac{\hat{A}_P(\omega_1,\Delta T)\hat{A}_P(\omega_2,\Delta T)\hat{A}_S^*(2\omega_P-\omega_S-\omega+\omega_1+\omega_2,0)}
{\left(
	\omega_P+\omega_1-\bar{\omega}_{ea}
	\right)
	\left(
	2\omega_P+\omega_1+\omega_2-\bar{\omega}_{ea}
	\right)
	\left(
	\omega-\bar{\omega}_{ea}
	\right)}
\end{multline}
where $\eta_{\hbox{\footnotesize{NVRB}}}= n_{\hbox{\footnotesize{NVRB}}} |\mu_{ea}|^4$, $n_{\hbox{\footnotesize{NVRB}}}$ is the number of scatterers involved in the NVRB process, and $\omega_{ea}$ is the energy of the electronic excited level involved in the NVRB process. Since the cross section of third-order nonlinear processes in graphene is enhanced by increasing the photon energy\cite{Wang2013146,concado2007}, we consider only the dominant case, i.e., $\tilde{\nu}_{ea}=2\tilde{\nu}_P$.

We describe the spectral FWM response assuming monochromatic fields with no inter-pulse delay:
$\hat{E}_{P}(\omega)=E_P\cdot \delta(\omega-\omega_P)$, $\hat{E}_{S}(\omega)=E_S\cdot \delta(\omega-\omega_S)$. From Eqs.\ref{eq:simulation},\ref{eq:simulationNVRB}, the CARS and NVRB nonlinear polarizations can be expressed as\cite{Mukamel1999,PotmaMukamel2013}:
\begin{equation}\label{eq:Pdelta_cars}
\begin{split}
P^{(3)}_{CARS}(\omega)
&\propto -\frac{\eta_{CARS} E_P^2 E_S^*}{(\omega_P-\bar{\omega}_{ba})(\omega-\omega_P-\bar{\omega}_{ca})(\omega-\bar{\omega}_{da})}=\\  &=\chi^{(3)}_{CARS}E_P^2 E_S^*
\end{split}
\end{equation}
\begin{equation}\label{eq:Pdelta_nvrb}
\begin{split}
P^{(3)}_{NVRB} (\omega) &
\propto -\frac{\eta_{NVRB}E_P^2 E_S^*}{(\omega_P-\bar{\omega}_{ea})(2\omega_P-\bar{\omega}_{ea})(\omega-\bar{\omega}_{ea})}=\\&=\chi^{(3)}_{NVRB}E_P^2 E_S^*
\end{split}
\end{equation}
which can be used to calculate the total FWM spectrum according to Eq.\ref{eq:FWMgeneral}. Fig.\ref{fig1_bis}a plots the electronically non-resonant case. The CARS polarization, defined by Eq.\ref{eq:Pdelta_cars}, is a complex quantity: the real part has a dispersive lineshape, while the imaginary part peaks at the phonon frequency $\omega_{ca}$. The NVRB polarization, defined by Eq.\ref{eq:Pdelta_nvrb}, is a positive real quantity. Accordingly, the FWM spectrum in the electronically non-resonant condition, $I(\omega_{as})^{NR}$, can be written as\cite{Mukamel1999}:
\begin{equation}
\begin{split}
I(\omega_{as})^{NR}&\sim \lvert
P^{(3)}_{NVRB}\rvert^2+
\lvert P^{(3)}_{CARS}\rvert^2+
2\Re(P^{(3)}_{NVRB})\Re(P^{(3)}_{CARS})\\
&\propto
\lvert\chi^{(3)}_{NVRB}\rvert^2+
\lvert\chi^{(3)}_{CARS}\rvert^2+
2\Re(\chi^{(3)}_{NVRB})\Re(\chi^{(3)}_{CARS})
\end{split}
\label{eq:FWMinterf}
\end{equation}
and it can be significantly distorted by the third term in Eq.\ref{eq:FWMinterf} depending on the relative weight of the two corresponding susceptibilities. The maximum of the signal, when the dispersive contribution is dominant, can be frequency shifted from the phonon frequency. This is the most common scenario, in which the dispersive lineshapes hampering a direct access to the vibrational characterization of the sample in terms of phonon frequencies and lifetimes. Such limitation is particularly severe when $\chi^{(3)}_{NRVB}$ is comparable to $\chi^{(3)}_{CARS}$ and the NVRB and CARS contributions have the same order of magnitude. This condition is common in the case of a weak vibrational resonant contribution ($\frac{\mu_{ba} \mu_{cb} \mu_{cd} \mu_{ad}}{|\mu_{ea}|^4}<<1$), as in the case of low concentrations of oscillators ($\frac{n_{\hbox{\footnotesize{CARS}}}}{n_{\hbox{\footnotesize{NVRB}}}}<<1$). Hence, this produces an intense NVRB signal and reduces the vibrational contrast, hindering the imaging of electronically non-resonant samples. This is the case for cells and tissues which need to be excited in the near infrared to avoid radiation damage\cite{Muller2007}.

For SLG, the linear dispersion of the massless Dirac Fermions makes the response always electronically resonant. In the case of FLG, absorption has a complex dependence on wavelength, as well as on the number of layers and their relative orientation, exhibiting, for instance, a tunable band gap in twisted bilayer graphene\cite{Zhang2009}. This is also reflected in the resonant nature of SR\cite{Wu2014,Wu2015}. However, approaching visible wavelengths, the absorption spectrum flattens above$\sim0.8$eV and it is $\sim (1-\pi e^2/2h)^N$ for Bernal-stacked $N$-layer graphene\cite{Mak2010}. Our exfoliated FLG are Bernal stacked, as also confirmed by the measured SR 2D peak shape in SR\cite{FerrariPRL2006, FerrariReview2013}. Accordingly, at the typical CARS wavelengths used here (784 and 894nm), SLG and Bernal FLG are electronically resonant, unlike  the situation for most biological samples\cite{Muller2007}. This results in an opposite sign in the CARS response, i.e. a spectral dip in $\Im(\chi^{(3)}_{CARS})$, related to two additional imaginary unit contributions in the denominator of Eq.\ref{eq:Pdelta_cars}, $(\omega_P-\bar{\omega}_{ba})$ and $(\omega-\bar{\omega}_{da})$, wherein the $i\gamma_{ba}$, $i\gamma_{da}$ components dominate. Further, the $-i$ contribution from $(2\omega_P-\bar{\omega}_{ea})$ results in a NVRB dominated by the imaginary part, as illustrated in Fig.\ref{fig1_bis}c,d,e.

Thus, the third term in Eq.\ref{eq:FWMinterf} must be replaced with the contribution from the interference of the spectral dip $\Im(\chi^{(3)}_{CARS})$ with the imaginary part $\Im(\chi^{(3)}_{NVRB})$. This leads to a signal that, under the electronically resonant regime, becomes\cite{Mukamel1999}:
\begin{equation} \label{eq:FWMinterf_Resonant}
\begin{split}
I(\omega_{as})^{R}
& \sim \lvert P^{(3)}_{NVRB}\rvert^2+
\lvert P^{(3)}_{CARS}\rvert^2+
2\Im(P^{(3)}_{NVRB})\Im(P^{(3)}_{CARS}) \\
&  \propto\lvert
\chi^{(3)}_{NVRB}\rvert^2+
\lvert\chi^{(3)}_{CARS}\rvert^2+
2\Im(\chi^{(3)}_{NVRB})\Im(\chi^{(3)}_{CARS})
\end{split}
\end{equation}
which indicates that the total FWM, at the phonon frequency, can be either a negative dip or a positive peak depending on the ratio between the vibrationally resonant and the non-resonant susceptibilities ($\chi^{(3)}_{CARS}/\chi^{(3)}_{NVRB}$), which depends only on the material under examination and not on the pulses used in the experiment. Such a qualitatively different interplay between NVRB and CARS, compared with the experimental lineshapes for $\Delta T=0$ in Fig.\ref{fig2}, unambiguously indicates the presence of electronic resonance in SLG and Bernal FLG. For a given material, the relative weight of the two FWM contributions can be in general modified by using pulsed excitation and tuning the temporal overlap between the PP and SP fields\cite{Volkmer2002}, i.e. changing $\Delta T$. The experimentally observed evolution of the FWM signal in FLG as a function of PP-SP delay in Fig.\ref{fig2} has a trend similar to that shown in Fig.\ref{fig1_bis}e,f as function of $\eta_{NRVB}/\eta_{CARS}$, validating the resonance-dominated scenario.

A more quantitative picture can be derived from Eqs.\ref{eq:simulation},\ref{eq:simulationNVRB}, where the PP and SP temporal profiles are taken into account, matching those retrieved from the experimentally measured autocorrelation (Fig.\ref{figsetup}).
\begin{figure*}
\centerline{\includegraphics[width=180mm]{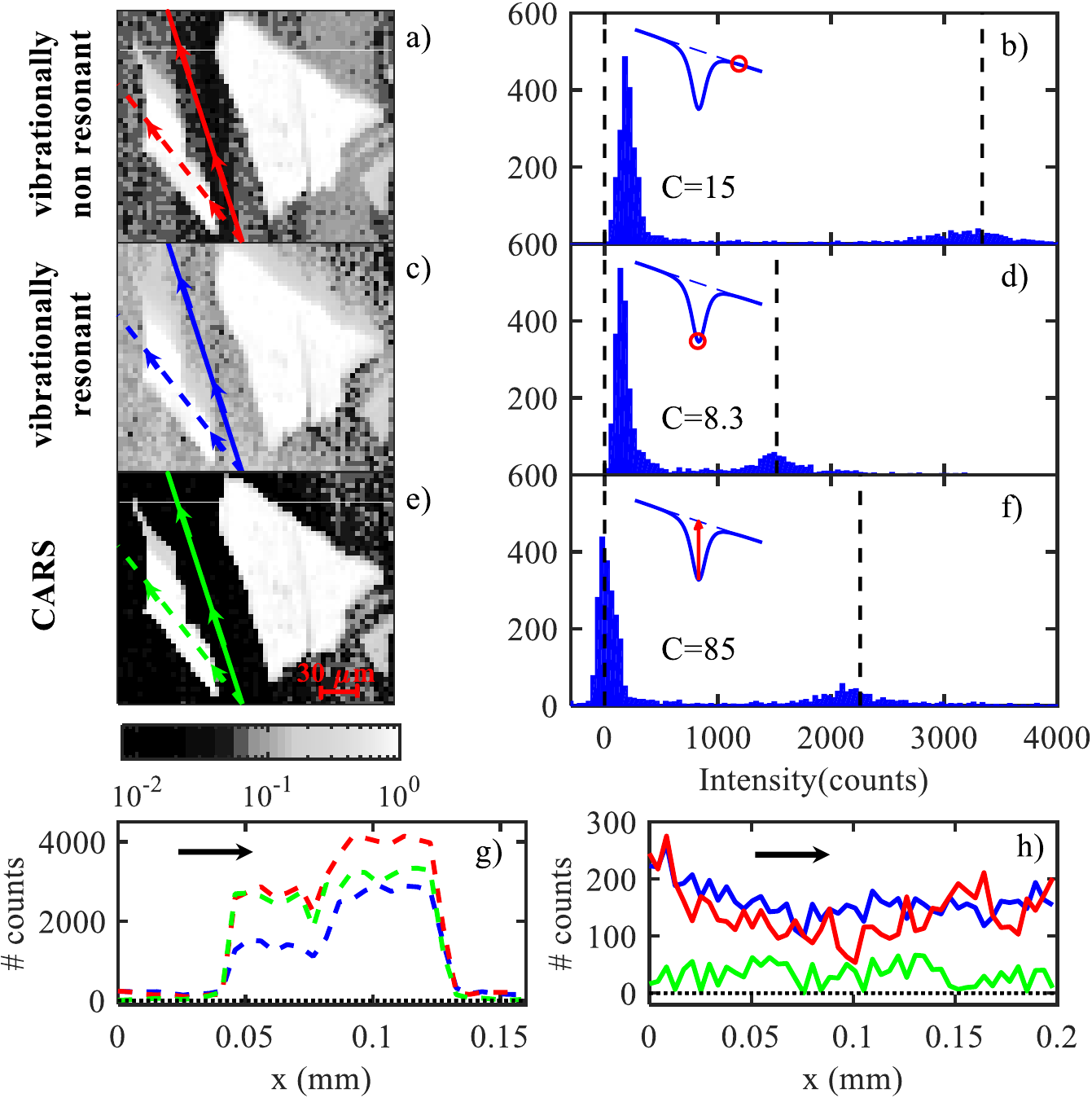}}
\caption{Nonlinear optical images of FLG measured under conditions of (a) a non-vibrationally resonant $\lambda_S$ at 891.5nm and (c) a resonant $\lambda_S$ at 894 nm and $\Delta T$=1.7 ps. e) CARS image of two FLG flakes, obtained by the spectral dip (see Eq.\ref{eq:imag_sottr}). b,d,f) Intensity histograms of a,c,e). The corresponding contrast $C$ is also reported. The black dashed lines represent the colormap boundaries of a,c,e). g,h) Intensity profiles along the scanning paths in and out of a FLG flake as highlighted in a,c,e  by dashed and full lines, respectively. }
\label{fig3}
\end{figure*}

As model parameters for the FLG we use the experimental SR value $\tilde{\nu}_{ca}$=1580cm$^{-1}$, with fitted $\tau_{ca}=1.1\pm 0.1$ps\cite{FerrariPRL2006,Bonini2007} (corresponding to FWHM(G)$=10$cm$^{-1}$) and $\tau_{da}=\tau_{ba}=\tau_{ea}=10 \pm 2$fs in agreement with the value measured for SLG\cite{Brida2013}. The ratio between NVRB and CARS contributions $\frac{\eta_{CARS}}{\eta_{NVRB}}=(3.0\pm0.7)\times10^{-5}$ is obtained by fitting to the experimental data in Fig.\ref{fig2} with Eqs.\ref{eq:FWMgeneral},\ref{eq:simulation2},\ref{eq:simulationNVRB2}. The resulting spectra (colored dashed lines in Fig.\ref{fig2}), evaluated by tuning only the PP-SP delays, are in good agreement with the experimental data, with $\frac{\eta_{CARS}}{\eta_{NVRB}}$ as the only adjustable parameter. This ratio, combined with Eqs.\ref{eq:Pdelta_cars},\ref{eq:Pdelta_nvrb}, allows us to extract the ratio between the third-order nonlinear susceptibilities for CARS and NVRB: $\frac{|\chi^{(3)}_{CARS}|}{|\chi^{(3)}_{NVRB}|}\sim 1.3$ at the G-phonon resonance.

The dependence of our spectral profiles on the inter-pulse delay, $\Delta T$, indicates that the peculiar FWM lineshapes for SLG and FLG originates from the interference between two electronically resonant radiation-matter interactions (NVRB and CARS) rather than from a matter-only Hamiltonian coupling the electronic continuum and a discrete phonon state (implying a resonance between the corresponding energies), resulting in the Fano resonance\cite{fano1961} suggested in Ref.\cite{LafetaArx2017}.

In the electronically non-resonant case, CARS provides access to the real part of $\chi^{(3)}$\cite{evans_coherent_2008}. However, due to the dispersive nature of the $\chi^{(3)}$ real part\cite{evans_coherent_2008}, it distorts the phonon lineshapes\cite{BoydBook}, unlike SR. In SLG and FLG the FWM signal arises from the imaginary (non dispersive) CARS susceptibility, and is amplified by its NVRB (third term in Eq.\ref{eq:FWMinterf_Resonant}). Thus, the signal can be used for vibrational imaging, unlike the non-resonant case\cite{evans_coherent_2008}.

The vibrationally resonant contribution $I$ can be isolated by subtracting from the $I_2$ FWM signal at $\tilde{\nu}_2-\tilde{\nu}_P\sim\tilde{\nu}_G$, the NRVB obtained by linear interpolation of the spectral intensities measured at the two frequencies at the opposite sides of vibrational resonance:
\begin{equation}
I=I_1-I_2+\frac{\tilde{\nu}_2-\tilde{\nu}_1}{\tilde{\nu}_3-\tilde{\nu}_1}(I_3-I_1)
\label{eq:imag_sottr}
\end{equation}
where the indexes i=1,2,3 refer to data at $\tilde{\nu}_1=\tilde{\nu}_P+1545$cm$^{-1}$, $\tilde{\nu}_2=\tilde{\nu}_P+1576$cm$^{-1}$, $\tilde{\nu}_3=\tilde{\nu}_P+1607$cm$^{-1}$ (i.e. with $\tilde{\nu}_2$ near to the G phonon frequency and $|\tilde{\nu}_{1,3}-\tilde{\nu}_{G}|$ greater than two half-widths at half maximum of the measured profiles, as shown in Fig.\ref{fig2}).

This combination of electronically resonant NVRB and CARS non-linear responses gives CARS images (i.e. retaining vibrational sensitivity) with signal intensities comparable to those of NVRB, for which sub-ms pixel dwell times have been demonstrated with the use of a point detector, e.g. photomultiplier\cite{Hendry2010}. In our case, the images in Fig.\ref{fig3} are obtained with a pixel dwell time$\sim200$ms using a Si-Charge-Coupled Device (CCD) array, aiming at a complete spectral characterization, and scanning the sample at fixed $\Delta T$ with stepper-motor translation stages.

Figs.\ref{fig3}(a--c) display nonlinear optical images measured at two different $\omega_S$, corresponding to vibrationally non-resonant and resonant conditions. Extracting for each image pixel $I_1$ (Fig.\ref{fig3}a), $I_2$ (Fig.\ref{fig3}c), and $I_3$, required for Eq.\ref{eq:imag_sottr}, we obtain an image with suppression of the signal not generated by FLG, as in Fig.\ref{fig3}e.

To obtain a quantitative comparison of the different images, we plot the pixel intensity histogram in Figs.\ref{fig3}b,d,f. This gives a bimodal distribution: one peak corresponds to the most intense pixels, associated with FLG ($I_{g}$) and the other is related to the weaker substrate signal ($I_{s}$). The ability to discriminate sample from substrate can be quantified in terms of 1) $I_{g}$ compared to $I_{s}$, evaluated as the difference $I_{g}-I_{s}$, and 2) the proximity of $I_{s}$ to $I=0$ in the histogram (dashed black line in Figs.\ref{fig3}b,d,f). These two features can be quantified by the contrast $C$ in order to compare the images: $C=\frac{\overline{I}_g-\overline{I}_s}{\overline{I}_s}$, where $\overline{I}_g$ and $\overline{I}_s$ are the mean FLG and substrate intensities, corresponding to the local maxima in the histograms in Figs.\ref{fig3}b,d,f. In Figs.\ref{fig3}g,h we plot the intensity profiles along two scanning paths, one inside (dashed) and the other adjacent to (full line) the FLG flake.

Comparing the three histograms (Figs.\ref{fig3}b,d,f), the vibrationally off resonant FWM image (NVRB only, Fig.\ref{fig3}b) has the highest $\overline{I}_g$. The visibility of the flakes is limited by the noise of the detector and by $\chi^{(3)}$ nonlinearity of the substrate. NVRB, lacking vibrational specificity, can also originate from the glass substrate outside the FLG flake ($\overline{I}_s\gg 0$), as indicated by the scanning profile in Fig.\ref{fig3}h (red line). This may become a critical limitation in those substrates with $\chi^{(3)}$ much larger than Si ($\lvert\chi^{(3)}\rvert$ $\sim2.5 \times 10^{-10}$ e.s.u.\cite{Potma2013}), such as Au ($\chi^{(3)}=4\times 10^{-9}$e.s.u\cite{BOYD201474, XENOGIANNOPOULOU2007217}). Similarly, the vibrationally resonant FWM, $I_2$, originating from concurrent CARS and NVRB processes (Fig.\ref{fig3}d), has a $\overline{I}_s\gg 0$ related to NRVB. The depth of the FWM dip (Fig.\ref{fig3}f) is related to the CARS signal intensity, and its vibrational sensitivity brings about a substantial contrast increase, as demonstrated by the close-to-zero average value of the (green) scanning profile in Fig.\ref{fig3}h.

In summary, by using an experimental time-delayed FWM scheme, CARS peaks equivalent to those seen in spontaneous Raman were obtained from graphene. By explaining the physical mechanism responsible for the FWM signal, we demonstrated that the spectral response can be described in terms of joint CARS and NVRB contributions concurring to the overall signal. Unlike non-resonant FWM, where dispersive lineshapes hamper vibrational imaging of biological systems, the resonant nature of FWM in graphene, which can be traced back to its peculiar electronic properties, mixes CARS and NVRB, resulting in Lorentzian profiles which are either peaks or dips depending on their relative strength. We also demonstrated that CARS can be used for vibrational imaging with contrast equivalent to spontaneous Raman microscopy and signal levels as large as those of the third order nonlinear response.
\section{acknowledgements}
We thank M. Polini for useful discussions. We acknowledge funding from the EU Graphene Flagship, ERC grant Hetero2D, EPSRC grants EP/L016087/1, EP/K01711X/1, EP/K017144/1. This project has received funding from the European Union’s Horizon 2020 research and innovation programme under grant agreement No. 785219 - GrapheneCore2.
\section{Methods}
The third -order response for the SLG and FLG samples can be obtained from the third-order polarization\cite{Mukamel1999}:
	\begin{multline}\label{3orderResponse}
	P^{(3)}(t)
	\propto
	N \int_0^{\infty}d\tau_3\int_0^{\infty}d\tau_2\int_0^{\infty}d\tau_1
	\E(t-\tau_3)\\
	\E(t-\tau_2 -\tau_3)\E(t-\tau_1-\tau_2-\tau_3) S^{(3)}(\tau_1,\tau_2,\tau_3)
	\end{multline}
where $N$ is the number of scatterers, $S^{(3)}(\tau_1,\tau_2,\tau_3)$ may be expressed as\cite{Mukamel1999}:
	\begin{multline}\label{S}
	S^{(3)}(\tau_1,\tau_2,\tau_3)
	\propto
	\left(i \right)^3 Tr\Big\{\mu(\tau_1+\tau_2+\tau_3)\\\big[\mu(\tau_1+\tau_2) ,\big[\mu(\tau_1),\big[\mu(0),\rho(-\infty) \big] \big]\big] \Big\}
	\end{multline}
and $\E(t)$ is the total electric field on the sample
\begin{equation}
\E(t)=\sum_{i=P,S} \big[ E_{i}(t,\Delta t_i)  +c.c. \big]=\sum_{i=P,S} \big[\hat{A}_{i}(t,\Delta_i t) e^{-i\omega_i t} +c.c.\big]
\end{equation}

Consider a SP at $\Delta t_S=0$ with $\Delta t=\Delta t_P$. The energy level diagrams in Fig.\ref{fig1}schematically illustrate the CARS and NVRB processes\cite{Mukamel1999}:
\begin{widetext}
	\begin{multline}\label{P_CARS}
	P^{(3)}_{CARS}(t)
	\propto
	\left(i\right)^3
	n_{\hbox{\footnotesize{CARS}}}\mu_{ba} \mu_{cb} \mu_{cd} \mu_{ad}
	\int_0^{\infty}d\tau_3\int_0^{\infty}d\tau_2\int_0^{\infty}d\tau_1
	{A}_P(t-\tau_1-\tau_2-\tau_3,\Delta t){A}_S^*(t-\tau_2 -\tau_3){A}_P(t-\tau_3,\Delta t)
	\\
	e^{-i\omega_P (t-\tau_1-\tau_2-\tau_3,\Delta t)}
	e^{+i\omega_S (t-\tau_2-\tau_3)}
	e^{-i\omega_P (t-\tau_3,\Delta t)}
	e^{-i\bar{\omega}_{ba} \tau_1}
	e^{-i\bar{\omega}_{ca} \tau_2}
	e^{-i\bar{\omega}_{da} \tau_3}
	\end{multline}
	\begin{multline}\label{P_NVRB}
	P^{(3)}_{NVRB}(t)	\propto
	\left(i\right)^3
	n_{\hbox{\footnotesize{NVRB}}} |\mu_{ea}|^4
	\int_0^{\infty}d\tau_3\int_0^{\infty}d\tau_2\int_0^{\infty}d\tau_1
	{A}_P(t-\tau_1-\tau_2-\tau_3){A}_P(t-\tau_2 -\tau_3){A}_S^*(t-\tau_3)
	\\
	e^{-i\omega_P (t-\tau_1-\tau_2-\tau_3,\Delta t)}
	e^{-i\omega_P (t-\tau_2-\tau_3,\Delta t)}
	e^{+i\omega_S (t-\tau_3)}
	e^{-i\bar{\omega}_{ea} \tau_1}
	e^{-i\bar{\omega}_{ea} \tau_2}
	e^{-i\bar{\omega}_{ea} \tau_3}
	\end{multline}
\end{widetext}
where $\bar{\omega}_{ij}=\omega_i-\omega_j -i\gamma_{ij}$.

By Fourier transform, the frequency dispersed signal can be expressed as:
\begin{equation}
P^{(3)}(\omega)=\int_{-\infty}^{\infty}P^{(3)}(t)e^{i\omega t} dt.
\end{equation}
In order to reduce the computational effort to calculate Eqs.\ref{P_CARS},\ref{P_NVRB}, we also write the pulse fields in terms of their Fourier transforms, obtaining:
	\begin{multline}\label{P_CARS_omega}
	P^{(3)}_{CARS}(\omega) \propto
	\eta_{\hbox{\footnotesize{CARS}}}
	\left(i\right)^3
	\int_{-\infty}^{\infty}dt \,e^{i\omega t}
	\int_0^{\infty}d\tau_3\int_0^{\infty}d\tau_2\int_0^{\infty}\\d\tau_1
	\int_{-\infty}^{\infty} d\omega_1
	\int_{-\infty}^{\infty} d\omega_2
	\int_{-\infty}^{\infty} d\omega_3
	\hat{A}_P(\omega_1,\Delta t)\hat{A}_S^*(\omega_2,0)\\\hat{A}_P(\omega_3,\Delta t)
	e^{-i(\omega_P+\omega_1)(t-\tau_1-\tau_2-\tau_3)}
	e^{+i(\omega_S+\omega_2)(t-\tau_2-\tau_3)}\\
	e^{-i(\omega_P+\omega_3)(t-\tau_3)}
	e^{-i\bar{\omega}_{ba} \tau_1}
	e^{-i\bar{\omega}_{ca} \tau_2}
	e^{-i\bar{\omega}_{da} \tau_3}
	\end{multline}
where $\eta_{\hbox{\footnotesize{CARS}}}= n_{\hbox{\footnotesize{CARS}}} \mu_{ba} \mu_{cb} \mu_{cd} \mu_{ad}$. In this way all the temporal integrals can be solved analytically:
\begin{widetext}
	\begin{multline}\label{P_CARS_omega_END-1}
	P^{(3)}_{CARS}(\omega) \propto
	-\eta_{\hbox{\footnotesize{CARS}}}
	\int_{-\infty}^{\infty} d\omega_1
	\int_{-\infty}^{\infty} d\omega_2
	\int_{-\infty}^{\infty} d\omega_3
	\delta(\omega-2\omega_P+\omega_S+\omega_1-\omega_2+\omega_3)\\
	\frac{\hat{A}_P(\omega_1,\Delta t)\hat{A}_S^*(\omega_2,0)\hat{A}_P(\omega_3,\Delta t)}{(\omega_P+\omega_1-\bar{\omega}_{ba})(\omega_P-\omega_S+\omega_1-\omega_2-\bar{\omega}_{ca})(
		2\omega_P-\omega_S+\omega_1-\omega_2+\omega_3-\bar{\omega}_{da})}
	\end{multline}
\end{widetext}
using the energy conservation, represented by the delta distribution:
\begin{equation}
\delta(\omega-2\omega_P+\omega_S-\omega_1+\omega_2-\omega_3)=\int_{-\infty}^{\infty}e^{i(\omega-2\omega_P+\omega_S-\omega_1+\omega_2-\omega_3)t},
\end{equation}
the $\omega_2$ integral can be simplified:
	\begin{multline}\label{eq:simulation2}
	P^{(3)}_{CARS}(\omega,\Delta t)
	\propto
	-\eta_{\hbox{\footnotesize{CARS}}}
	\int_{-\infty}^{\infty}d\omega_1
	\int_{-\infty}^{\infty}d\omega_3\\
	\frac{\hat{A}_P(\omega_3,\Delta t)\hat{A}_P(\omega_1,\Delta t)\hat{A}_S^*(2\omega_P-\omega_S-\omega+\omega_3+\omega_1,0)}
	{\left(
		\omega_P+\omega_3-\bar{\omega}_{ba}
		\right)
		\left(
		\omega-\omega_P-\omega_1-\bar{\omega}_{ca}
		\right)
		\left(
		\omega-\bar{\omega}_{da}
		\right)}
	\end{multline}
In a similar way, using  $\eta_{\hbox{\footnotesize{NVRB}}}=n_{\hbox{\footnotesize{NVRB}}} |\mu_{ea}|^4$, Eq.\ref{P_NVRB} can be written as:
	\begin{multline}\label{eq:simulationNVRB2}
	P^{(3)}_{NVRB}(\omega,\Delta t)
	\propto
	-\eta_{\hbox{\footnotesize{NVRB}}}
	\int_{-\infty}^{\infty}d\omega_1
	\int_{-\infty}^{\infty}d\omega_2\\
	\frac{\hat{A}_P(\omega_1,\Delta t)\hat{A}_P(\omega_2,\Delta t)\hat{A}_S^*(2\omega_P-\omega_S-\omega+\omega_1+\omega_2,0)}
	{\left(
		\omega_P+\omega_1-\bar{\omega}_{ea}
		\right)
		\left(
		2\omega_P+\omega_1+\omega_2-\bar{\omega}_{ea}
		\right)
		\left(
		\omega-\bar{\omega}_{ea}
		\right)}
	\end{multline}


\begin{thebibliography}{99}

	\bibitem{Soavi2018} G. Soavi et al. Nat. Nanotechnol., \textbf{15}, 1748 (2018)
	
	\bibitem{Hendry2010} E. Hendry et al. Phys. Rev. Lett. \textbf{105}, 097401 (2010)
	
	\bibitem{BoydBook}  R. W. Boyd, Nonlinear Optics,  (Academic Press, 2008)
	
	\bibitem{PotmaMukamel2013} E. O. Potma and S. Mukamel, in \emph{Coherent Raman Scattering Microscopy}, edited by J. Cheng and X. S. Xie (CRC Press, 2012)

	\bibitem{Khurgin2014} J.B. Khurgin, Appl. Phys. Lett. \textbf{104}, 161116 (2014)
	
	\bibitem{Wu2011} R. Wu et al. Nano Lett. \textbf{11}, 5159 (2011)
	
	\bibitem{Gu2012} T. Gu et al. Nat. Photonics \textbf{6}, 554 (2012)
	
	
	\bibitem{duong_probing_2012} D.L. Duong et al. Nature \textbf{490}, 235 (2012)
	
	\bibitem{FerrariPRL2006} A.C. Ferrari et al. Phys. Rev. Lett. \textbf{97}, 187401 (2006)
	
	\bibitem{FerrariReview2013} A.C. Ferrari, D.M. Basko, Nat. Nanotechnol. \textbf{8}, 235 (2013).
	
	\bibitem{Begley1974} R.F. Begley et al. Appl. Phys. Lett. \textbf{25}, (1974).
	
	\bibitem{Duncan1982} M.D. Duncan et al. Opt. Lett. \textbf{7}, 350 (1982)
	
	\bibitem{Zumbusch1999} A. Zumbusch et al. Phys. Rev. Lett. \textbf{82}, 4142 (1999)
	
	\bibitem{Cheng2002} J.X. Cheng et al. J. Opt. Soc. Am. B \textbf{19}, 1363 (2002)
	
	\bibitem{Mukamel1999} S. Mukamel, Principles of Nonlinear Optical Spectroscopy,  (Oxford University Press, New York, 1999)
	
	\bibitem{Cui2009} M. Cui et al. Opt. Lett. \textbf{34}, 773 (2009)
	
	\bibitem{Pestov265} D. Pestov, et al. Science \textbf{316}, 265 (2007)
	
	\bibitem{evans_coherent_2008} C.L. Evans, X.S. Xie, Annu. Rev. Anal. Chem. \textbf{1}, 883 (2008)
	
	\bibitem{Evans16807} C. L. Evans et al. Proc. Natl. Acad. Sci. U.S.A \textbf{102}, 16807 (2005)
	
	\bibitem{CiceroneNat2015}  C. H. Camp Jr and M. T. Cicerone, Nat. Photonics, \textbf{5}, 295 (2015)
	
	\bibitem{Lee2012} Y. J. Lee et al. ACS Macro Lett. \textbf{1} (11), 1347 (2012)
	
	\bibitem{Lim2006} S. H. Lim et al. J. Phys. Chem. B \textbf{110} (11), 5196 (2006)
	
	
	\bibitem{Lim2011} R.S. Lim, et al. J. Lipid Res. \textbf{52}  2177 (2011)
	
	\bibitem{Steuwe2011} C. Steuwe et al. Nano Lett. \textbf{11}, 5339 (2011)
	
	\bibitem{Ikeda2009} K. Ikeda, K. Uosaki Nano Lett. \textbf{9}, 1378 (2009)
	
	\bibitem{Lee2010} Y.J. Lee et al. Phys. Rev. B \textbf{82}, 165432 (2010)
	
	\bibitem{Dovbeshko2010} G. Dovbeshko et al. Nanoscale Res. Lett. \textbf{9}, 263 (2014)
	
	\bibitem{Baldacchini2010} T. Baldacchini, R. Zadoyan, Opt. Express \textbf{18}, 19219  (2010)
	
	\bibitem{Potma2013} J. Brocious, E.O. Potma, Mater. Today \textbf{16}, 344  (2013)
	
	\bibitem{ferrante_nc} C. Ferrante et al. Nat. Commun., \textbf{9}, 308 (2018)
	
	\bibitem{TRCARS2017} J. Koivistoinen et al. J. Phys. Chem. Lett., \textbf{8}, 4108 (2017)
	
	\bibitem{LafetaArx2017} L. Lafeta et al. Nano Lett., 17 (6), pp 3447–3451 (2017)
	
	
	\bibitem{LiS2009} X. S. Li, W. W. Cai, J. H. An, S. Kim, J. Nah, D. X. Yang, R. Piner, A. Velamakanni, I. Jung, E. Tutuc, S. K. Banerjee, L. Colombo, R. S. Ruoff, \textit{Science} \textbf {324}, 1312 (2009).
	
	\bibitem{BaeNN2010} S. Bae, H. Kim, Y. Lee, X. Xu, J. S. Park, Y. Zheng, J. Balakrishnan, T. Lei, H. R. Kim, Y. I. Song, Y. J. Kim, K. S. Kim, B. Ozyilmaz, J. H. Ahn, B. H. Hong, S. Iijima, \textit{Nat. Nanotechnol.} \textbf {5}, 574 (2010).
	
	\bibitem{BonaMT2012} F. Bonaccorso, A. Lombardo, T. Hasan, Z. P. Sun, L. Colombo, A. C. Ferrari, \textit{Mater. Today} \textbf {15}, 564 (2012).
	
	\bibitem{BaskPRB2009} D. M. Basko, S. Piscanec, A. C. Ferrari, \textit{Phys. Rev. B} \textbf {80}, (2009).
	
	\bibitem{DasNN2008} A. Das, S. Pisana, B. Chakraborty, S. Piscanec, S. K. Saha, U. V. Waghmare, K. S. Novoselov, H. R. Krishnamurthy, A. K. Geim, A. C. Ferrari, A. K. Sood, \textit{Nat. Nanotechnol.} \textbf {3}, 210 (2008).
	
		
	
	\bibitem{Novoselov2005PNAS} K. S. Novoselov et al. Proc. Natl. Acad. Sci. USA \textbf{102} (30), 10451 (2008)
	
	\bibitem{Krauss:09} G. Krauss et al. Opt. Lett., \textbf{34}, 2847 (2009)
	
	\bibitem{Marangoni:09} M. Marangoni et al. Opt. Lett., \textbf{34}, 3262 (2009)
	
	\bibitem{Bonini2007} N. Bonini et al. Phys. Rev. Lett. \textbf{99}, 176802 (2007)
	
	\bibitem{Tomadin2013} A. Tomadin, et al. Phys. Rev. B \textbf{88}, 035430 (2013)
	
	\bibitem{Brida2013} D. Brida et al. Nat. Commun. \textbf{4}, 1987 (2013)
	
	\bibitem{Volkmer2002} A. Volkmer et al. Appl. Phys. Lett. \textbf{80}, (2002)
	
	\bibitem{Kamga:80} F.M. Kamga, M.G. Sceats, Opt. Lett. \textbf{5}, 126 (1980)
	
	\bibitem{Sidorov-Biryukov:06} D.A. Sidorov-Biryukov, E.E.Serebryannikov, A.M. Zheltikov, Opt. Lett. \textbf{31}, 2323 (2006)
	
	\bibitem{Selm:10} R. Selm, et al. Opt. Lett. \textbf{35}, 3282 (2010).
	
	\bibitem{Kumar:11} V. Kumar et al. Opt. Express \textbf{19}, 15143 (2011).
	
	\bibitem{Batignani2016} G. Batignani et al. Sci. Rep. \textbf{6} (2016)
	
	\bibitem{Wang2013146} P. Wang et al. Chem. Phys. Lett. \textbf{556}, 146 (2013)
	
	\bibitem{concado2007} L.G. Can\c{c}ado, et al. Phys. Rev. B \textbf{76}, 064304 (2007)
	
	\bibitem{Muller2007} M. Muller, A. Zumbusch, ChemPhysChem \textbf{8}, 2156 (2007)
	
	\bibitem{Zhang2009} Y. Zhang, et al. Nature \textbf{459}, 820 (2009)
	
	\bibitem{Wu2014} J. B. Wu, et al. Nat. Commun. \textbf{5}, 5309 (2014)
	
	\bibitem{Wu2015} J. B. Wu, et al. ACS Nano \textbf{9}, 7, 7440 (2015)
	
	\bibitem{Mak2010} K. F. Mak, et al. Proc. Natl. Acad. Sci. USA \textbf{107}, 14999 (2010)	
		

	\bibitem{fano1961} U. Fano, Phys. Rev. \textbf{124}, 1866 (1961)
	
	\bibitem{BOYD201474} R. W. Boyd et al. Opt. Commun., \textbf{326}, 74 (2014)
	
	\bibitem{XENOGIANNOPOULOU2007217} E. Xenogiannopoulou et al. Opt. Commun., \textbf{275}, 217 (2007)
	
\end{thebibliography}
\end{document}